\documentclass[12pt]{article}

\usepackage{amsmath,amssymb,amsthm,graphicx,latexsym,subfigure}

\def\v{{\bf v}}
\def\bnabla{\mbox{\bf$\nabla$}}
\def\f{{\bf f}}

\begin{document}
	\begin{center}
		{\bf Analogue White Hole Horizon and its Impact on Sediment Transport}
	\end{center}
	\vspace{0.5cm}
	\begin{center}
		Debasmita Chatterjee \footnote{E-mail: $debasmitachatterjee424@gmail.com$},
		Praloy Das \footnote{E-mail: $praloydasdurgapur@gmail.com$},
		Subir Ghosh \footnote{E-mail: $subir_{-}ghosh2@rediffmail.com$}\\
		and\\
		B. S. Mazumder \footnote{E-mail: $ bsmazumder@gmail.com$}\\
		\vspace{1cm}
		Physics and Applied Mathematics Unit, Indian Statistical
		Institute\\
		203 B. T. Road, Kolkata 700108, India \\
	\end{center}
	\vspace{1cm}
	{\textbf{Abstract:~~}}\\
	Motivated by the ideas of analogue gravity, we have performed experiments in a flume where an analogue White Hole horizon is generated, in the form of a wave blocking region,  by suitably tuned uniform fluid (water) flow and counter-propagating shallow water waves. We corroborate earlier experimental observations by finding a critical wave frequency for a particular discharge above which the waves are effectively blocked beyond the horizon. An obstacle, in the form of a bottom wave, is introduced to generate a sharp blocking zone. All previous researchers used this obstacle.
	
	A novel part of our experiment is where we do not introduce the obstacle and find that wave blocking still takes place, albeit in a more diffused zone. Lastly we replace the fixed bottom wave obstacle by a movable sand bed to study the sediment transport and the impact of the horizon or wave blocking phenomenon on the sediment profile. We find signatures of the wave blocking zone in the ripple pattern.
   \newpage

{\bf{INTRODUCTION:}}\\
Ever since Hawking's theoretical discovery \cite{haw} that Black Holes can radiate and Bekenstein's proposal \cite{bek} of identifying the horizon area of a Black Hole with its entropy, there has been a paradigm shift in attempts to understanding quantization of gravity. In the framework of Jacobson \cite{jac} the above identification is put in a robust form where Einstein equations appear as equations of state, as in classical thermodynamics. Following Padmanabhan \cite{pad} and Verlinde \cite{ver} it now seems natural to understand gravity as an emergent phenomenon in a macroscopic sense. Gravity ceases to be a fundamental interaction like Electromagnetism and Weak and Strong interactions and it seems futile to try to gravity in the conventional way.

However,  all of the above ideas rest on the physical existence of Hawking radiation and so far it has not been possible to observe it directly for the simple reason that for a typical astrophysical Black Hole the Hawking temperature is extremely tiny, (in fact numerically less than the CMB temperature). This has led researchers to look for experimentally accessible analogue systems where effective Black Hole metric can be simulated and dynamics of the relevant excitations in the effective Black Hole metric can be directly studied. The hope is to directly observe signatures of analogue Hawking radiation.

There appear to be several analogue systems where different aspects of the Hawking radiation phenomenon can be studied; optical systems \cite{opt} and the related Metamaterial scenario \cite{mat}, Bose condensate \cite{bose},  fluid system \cite{unruh}, in a specifically astrophysical use of the acoustic analogy \cite{astro}. (For a comprehensive review with references see \cite{rev,rev1}.) In the present paper we will focus on the analogue fluid model \cite{unruh}. Unruh \cite{unruh} had pioneered the idea that a wave-like disturbance impressed on a flowing fluid with non-uniform velocity is structurally same as the dynamics of an excitation in an effective curved spacetime. Hence a proper choice of flow and impressed wave  can simulate a Black Hole-like effective metric leading to the formation of an analogue event horizon if the wave tries to penetrate an opposing background fluid flow that has a velocity larger than the effective wave velocity. This idea was experimentally tested by Rousseaux et.al \cite{rou} and by Weinfurtner et.al. \cite{wein} in a flume. An obstacle was placed on the bottom of the flume and it was seen that water waves could not overcome the opposing flow at a certain position above the obstacle. Hence a horizon was created beyond which the waves were absent and only the background flow of fluid remained. From a purely classical fluid dynamics context this phenomenon is quite common and observable in various natural settings \cite{rev1}. A river mouth ending in the sea is an example of a natural white hole where the the river flow blocks the sea waves. The circular jump in the kitchen sink is a more controlled example. Another interesting example is  the whale
fluke-print {\footnote{ As a whale swims or dives, it releases a vortex ring behind its
fluke at each oscillation. The flow induced on the free surface is directed radially and forms a oval patch that gravity waves cannot enter whereas capillary waves are seen on its boundary.}}. Indeed, it needs to be pointed out that the above phenomena are analogues of White Holes, the time reversed Black Hole, that is also a solution of the Einstein equation. In fact laboratory analogues, so far studied, are always analogue White Holes. So the novelty of the recent theoretical and experimental works lies in the interpretation of the results in the context of Analogue Hawking effect. The analysis of \cite{wein} showed that the waveform near the horizon consisted of the incoming long wavelength wave (that was impressed externally), and two short wavelength waves being swept back with the flow. Furthermore the amplitudes of the latter appeared in the exponential ratio as is true for Hawking radiation, that is predicted to be black body like. But most interestingly, one of the short wavelength waves had a negative frequency in the co-moving frame \cite{rou,wein} indicating directly the presence of mode mixing \cite{fluid}. The interpretation of the experimental observation that  waves are absent beyond the horizon is explained by noting that there is mode mixing at the horizon. The incoming long wavelength mode is converted to two short wavelength modes that are swept away with the flow and can not travel beyond the horizon.

 So far experimental studies in analogue fluid models in the context of Hawking effect have been confined to the study of the water surface profile. The real space water surface is  Fourier transformed  to analyze the contribution of various modes and their individual strengths.  Subsequently  the thermal nature of the Hawking radiation is verified. In the analogue scenario, this reduces to the study of  conversion incoming shallow water waves to deep water waves at the wave blocking zone (or analogue horizon) that are swept along the flow. (For a discussion on water waves see eg. \cite{dean}.) We, in the present experimental work, consider  a rectangular channel or flume  in which oppositely moving uniform flow of water and imposed wave-like disturbances are superimposed to produce the horizon (or blocking zone) beyond which the wave is absent. The schematic diagram of the experimental setup is presented in Figure 1. Here AB and CD represent the total flume and obstacle respectively and the uniform water flow is from A towards B. The actual wave blocking phenomenon in our flume is shown in  Figure 2. We   have observed the existence of a critical frequency of the incoming shallow water wave for a particular discharge, above which frequency the incoming wave is effectively blocked. This agrees with previous observations \cite{rou1}. In these experimental setups the bottom of the flume consists of a fixed obstacle, referred to here as the bottom wave, (see Figure 1), that helps the wave blocking in two ways: on the one hand it increases the opposing background flow velocity at the top of the bottom wave and on the other hand it decreases the velocity of the incoming shallow water since the latter decreases as the water depth decreases. This allows the formation of a sharply defined blocking region (see Figure 2).

 In our experiment, apart from the above, we have  also ventured in an uncharted path: effect of wave blocking on sediment bed. This requires an important change in the experimental setup that is the bottom wave obstacle is removed and instead the flat bottom of the flume is covered with sand that acts as the sediment bed. In absence of the bottom wave the blocking region becomes slightly diffuse although it is still clearly visible (see Figure 3 where effect of wave blocking in flume without the bottom wave obstacle or sediment is presented). Figure 4 shows a typical smooth  bed with fine  sediment particles before starting of an experiment. For ripples to appear in the sediment bed the flow discharge or effectively  the fluid velocity has to be chosen in
 such a way that the bottom shear stress is slightly  higher than the threshold value for the
 initiation of sediment movement at the undisturbed plane sand bed. We show the uniform ripple structure for a steady flow in Figure 5.   We compare  the bed profile where uniform ripples are formed by the uniform flow only (see Figure 5) with the bed profile where both flow and opposing wave interact to create the blocking zone (see Figure 6). We have found signatures of the blocking region on the sediment bed profile as demonstrated in Figure 7. Note that photograph in Figure 7 consists of about ten  photographs of consecutive portions of the sediment bed which were later edited to combine together to make one single photograph.  Figures 6 and 7 are the new and most important observations of the present work.

 The paper consists of the following sections: In Section II we provide a brief outline of the mathematics involved that leads to the analogue gravity scenario in fluid mechanics. Section III describes the experimental setup. Section IV consists of our experimental findings on the presence of a critical frequency above which the incoming waves are effectively blocked.  Here the flume includes the bottom wave obstacle and is without sediment. Section V is devoted to the new setup where we remove the bottom wave obstacle and instead include the sediment in the form of sand bed. The paper ends in Section VI with our conclusions  and future prospects  of the present work.\\

 {\bf{Section II: Mathematical framework of analogue gravity in fluid}}\\

The basic mathematical framework is quite straightforward (for details see \cite{rev1}). The Euler equation for the non-dissipative fluid with velocity field ${\v}$ is
\begin{equation}
\frac{d{\v}}{dt}=\dot{\v}+(\v.\bnabla)\v=-\frac{\bnabla p}{\rho}+ {\bf g} +\frac{\f}{\rho}
\label{vel}
\end{equation}
where $p$ is the pressure, ${\bf g}=-g\vec{e_z}$ the  acceleration due to gravity and $\f=-\rho \bnabla_\parallel V^\parallel$ a horizontal and irrotational force in the $x$ direction ($\parallel$) driven by the potential $V^\parallel$ which is at the origin of the flow.\\

 We assume the flow to be without   vorticity  and hence $\bnabla \times \v =0$ . We find   $(\v.\bnabla)\v=(\bnabla \times \v )\times \v+\frac{1}{2}\bnabla(v^2)=\frac{1}{2}\bnabla(v^2)$ with $\v=\bnabla\phi$ where $\phi$ is  the velocity potential. Without loss of generality the background flow $\v_B$ is taken to be   stationary, irrotational and horizontal: $\bnabla_\perp \v_B=0$, $\v_B=\v^\parallel_B \rightarrow \bnabla_\parallel .\v_B=0$. A velocity perturbation $\delta v$ of the background flow $\v_B$ with a corresponding vertical displacement $\delta h$ is considered. For a curl free  velocity perturbation $\delta v$ we  define a corresponding perturbed velocity potential $\delta \phi$. We skip the details of deriving the boundary condition for the perturbing potential $\delta \phi $ \cite{rev1} and introduce a Taylor series expansion for  $\delta\phi$:
\begin{equation}
\delta\phi(x,y,z)=\sum\limits_{n=0}^{\infty}\frac{z^n}{n!}\delta\phi_{(n)}(x,y).
\label{6}
\end{equation}

Following \cite{unruh,rev1} for the case of the wavelengths much smaller than the water depth, one can develop a wave equation for $\delta \phi_{(0)}$,
\begin{equation}
\partial_t^2\delta\phi_{(0)}+2(\v^\parallel_B.\bnabla_\parallel)\partial_t\delta \phi_{(0)}+(\v_B^\parallel\otimes\v_B^\parallel-gh)\bnabla^2\delta\phi_{(0)}=0.
\end{equation}
The above can be expressed as a  Beltrami-Laplace equation,
\begin{equation}
\Box \delta \phi_{(0)}=\frac{1}{\sqrt{-g}}\partial_\mu(\sqrt{-g}g^{\mu\nu}\partial_\nu \delta \phi_{(0)})=0
\end{equation}
 with the  identification of $g^{\mu\nu}$ as an inverse metric given by,
\begin{equation}
g^{\mu\nu}=
\begin{pmatrix}
1&\vdots&\v^\parallel_B\\
\ldots\ldots\ldots&.&\ldots\ldots\ldots\ldots\\
\v^\parallel_B&\vdots&\v^{\parallel 2}_B-ghI
\end{pmatrix}
\end{equation}

 Again, defining $g^{\mu\nu}g_{\mu\sigma}=\delta^\nu_\sigma$ we get the  metric $g_{\mu\nu}$, popularly referred to as the acoustic metric in the conventional Painlev\'e-Gullstrand form,
\begin{equation}
g_{\mu\nu}=\frac{1}{c^2}
\begin{pmatrix}
gh-\v^{\parallel 2}_B&\vdots&\v^\parallel_B\\
\ldots\ldots\ldots&.&\ldots\ldots\ldots\ldots\\
\v^\parallel_B&\vdots&-1
\end{pmatrix}
\end{equation}
where $c=\sqrt{gh}$, the velocity of water waves in shallow water, plays the role of  the analogue of  velocity of light. Thus the acoustic metric has a singularity when $\v^{\parallel }_B=\sqrt{gh}$ that is the flow velocity becomes equal to the wave velocity but in opposite direction to it.

The dispersion relation is obtained by substituting a plane wave disturbance of the form $\delta \phi _{(0)}\sim exp (i\vec k.\vec x -i\omega t)$. One finds
 \begin{equation}
(\omega -\vec k.\vec v_b)^2=ghk^2.
\label{dis}
\end{equation}
This is infact the shallow water wave limit of the full dispersion relation
\begin{equation}
(\omega -\vec k.\vec v_B)^2=gk tanh (kh).
\label{fdis}
\end{equation}
Let us now come to the flume experiments.\\
\newpage

\vspace {.3cm}
{\bf{Section III: Experimental setup}}\\
\vspace {.5cm}

{\it{The test channel or Flume}}: The experiments were carried out in a specially designed recirculating
flume (Figure 1) at
the Fluvial Mechanics Laboratory, Physics and Applied Mathematics Unit, Indian Statistical
Institute, Kolkata. Both experimental and recirculating
channels of the flume are of the same
dimensions (10 m long, 0.5 m wide and 0.5 m deep). The experimental walls of the flume are
made of Perspex windows with a length of 8 m, providing a clear view of the flow. One
centrifugal pump providing the flow is located outside the main body of the flume. The inlet
and outlet pipes are freely suspended from the overhead structure. The outlet pipe is fitted
with one bypass
pipe and a valve, so that by adjusting the valve in the outlet, the flow can be
controlled at a desired speed. The upstream end of the channel is divided into three subchannels
of equal dimensions, and one honeycomb cage is placed at each end of the subchannels
to ensure smooth, vortexfree
uniform water flow through the experimental channel.
An electromagnetic discharge meter with a digital display is fitted with the outlet pipe to
facilitate the continuous monitoring of the discharge. Water depth is kept constant at a depth of  25
cm for all experiments and the hydraulic slope shows to be of the order of 0.0001.
Thereafter, one obstacle like waveform structure made of smooth Perspex is fabricated
in the laboratory and placed at the bottom of the flume for experiment. The angles of the
stoss-side
and lee-side
slopes of the structure are approximately $35^0$ and $6^0$ respectively. The
structure  is 2.28 m long with a  crest height of 15 cm and spans the entire width of
the flume. The obstacle is painted with epoxy paint to make the surface smooth. Particular
care is taken to design the obstacle to minimize or avoid the flow separation. Our obstacle is
scaled with the obstacle of the experiment in \cite{rou1}.\\
{\it{Wavemaker}}:
A piston type
wavemaker
is mounted at the downstream end of the flume to generate
surface waves against the current (Figure 1). The wavemaker
is fabricated in the institute
workshop. Two six-inch
wheels are fitted at the end of a spindle, which has got a gear in its
middle position. One crank and shaft is connected at the rim end of each wheel. The shafts
are allowed to pass through a guide to restrict their motion in vertical direction only. A
triangular shaped six inch
cylinder (closed at both ends) is fitted at the other ends of the
shafts. When the spindle is rotated with the help of motor, the cylinder moves to and fro in
the vertical direction. The wavemaker
is placed in such a way that the triangular cylinder
remained partially submerged in water when it is at its extreme positions: topmost and
lowermost. This is done to avoid the generation of small unwanted waves and disturbances in
the flow. The amplitude of oscillations is maintained at 11 cm. Oscillations are produced at a
right angle to the steady unidirectional current, which leads to the surface waves propagating
against the current. The wavemaker
is fixed with a Variac (variable resistor) to control the
frequency of oscillation. Calibration is made using tachometer for frequency variation. The
coordinate system of the measurement is as follows: x positive downstream and z positive
upward. As the flume used in the present study is a recirculating
one, the generated surface
waves may get recirculated
with the flow, which may bring more complexity in the flow. To
avoid this complexity, a wave absorber
is placed further downstream behind the wavemaker.
Here, the experiments are performed over a range of discharges and frequencies of
upstream waves against the flow over the obstacle. For each flow discharge i.e. flow velocity
over the obstacle, a critical frequency of counter-propagating
shallow water waves generated
from the wavemaker
is observed for the occurrence of wave blocking
on the lee side of the
obstacle, that means, the effective white hole horizon is experimentally observed for a
particular frequency under a given flow discharge. The conversion from the shallow water
waves generated from the wavemaker
with a frequency over the obstacle to the deep water
waves occurs, when a counter current
becomes sufficiently strong to block the upstream
propagation of shallow water waves.

\vspace {.3cm}
{\bf{Section IV: Analogue White Hole and partial wave blocking below critical frequency}}\\
\vspace {.5cm}

In this section we provide the first set of observations of wave blocking or equivalently generation of the analogue White Hole horizon in experiments performed in our flume. Our aim is to show that for a given discharge, there exists a critical frequency above which imposed waves are blocked by the horizon. The blocking zone (or horizon) appears close to the crest of the obstacle, between the crest (inside positions C and D of Figure 1)  and the wavemaker (position B in Figure 1). In order to observe the passage of waves beyond the horizon we fix our attention on the region close to the crest of the obstacle but in between positions A and C of Figure 1. We perform the experiment by keeping the discharge fixed and we increase the frequency of the wavemaker until there are effectively no fluctuations of the water surface in the observation point. There is no change if we increase the frequency of wavemaker still further. This indicates that for a given discharge a critical frequency exists above which the passage of waves over the obstacle are not allowed. However waves having frequency below the critical frequency can pass over the obstacle. This experiment is repeated for a number of discharges having different critical frequencies. From Figure 9 we also observe that the value of critical frequency decreases as the discharge increases.

 The images of the amplitudes of a given
frequency of counter propagating waves traveling over the obstacle were recorded by a
camera. As all the data from the camera were in terms of pixel units, we converted those data
to metric units. The recorded images of amplitudes were analyzed for a range of frequencies
and discharges using  digital imaging technique for better realization.   Data were recorded for the variation of blocking
frequencies with corresponding discharges.

 Figure 8 shows the
 plots of wave amplitudes against the frequency for different discharges Q = $0.0141 m^3/sec$,~$0.0166 m^3/sec$ and $0.0194 m^3/sec$. As summarized in Table 1, notice that the
 blocking frequency  decreases   as the discharge increases. Furthermore, variations of amplitude of the wave
with frequency for three different discharges can also be seen in Table 1. The amplitudes generically increase with increasing discharge.
 Our observations regarding critical frequency agree with earlier findings \cite{rou1}.

\vspace {.3cm}
{\bf{Section V: Sediment transport and effect of wave blocking on bed profile }}\\
\vspace {.5cm}

An intensely active area of fluid dynamics research is sediment transport. In general when fluid bed is deformable the combined fluid-bed system results in a complicated non-linearly coupled system. The fluid flow directly affects the bed form and the latter, as a back reaction, can modify fluid flow. The study of sediment transport has wide practical applications in rivers and oceans  and equally important is the study in laboratory flumes in a controlled environment. Studies of sediment transport from a modern perspective can be found in \cite{newbook}.

In laboratory flumes one generally  performs experiments
with initially smooth and flat sand beds, as in Figure 4. Bed forms with regular ripples and dunes are
developed when sediment transport starts. The flow discharge or effectively  the fluid velocity has to be chosen in
such a way that the bottom shear stress is slightly  higher than the threshold value for the
initiation of sediment movement at the undisturbed plane sand bed, i.e. when there is no sand
transport at the bed. Hjulstrom \cite{hij} suggested that there exists for each sand grain size a
certain velocity, called critical velocity, above which it will experience erosion on the
sediment bed. Subsequently, for initiation of sediment particle movement Shields \cite{sh}
 introduced a dimensionless quantity called shear stress, in terms of  various relevant physical parameters of the system such as acceleration due to gravity, densities of the fluid and sediment, kinematic viscosity, etc.  When shear stress exceeds the critical shear stress, the sediment motion begins to set in. As fluid flows over a flat sediment bed in a flume, the bed forms
are generated along the flume surface. For example, Figure 5 shows photographs of the bed
forms with small asymmetric ripples in an open channel flow. The bedform dimensions, such
as amplitudes and wavelengths, depend on the flow velocity and the bed characteristics. This
process seems to reach a steady state over a period of time and at each maximum velocity,
the 'fully developed' bedform (equilibrium category) seems to be evolved from an
approximately steady uniform flow.

Interesting contributions along these lines
were reported by Allen \cite{all}, where a very detailed account on bed forms and related
features from a geologist/sedimentologists
point of view is available. The most extensive
experimental studies on bed forms were made by US Geological Survey at the Colorado State
University (CSU). A summary of the data of all the flume experiments was presented by Guy et.al. \cite{guy}.  Since then, several experimental
investigations were performed under controlled conditions in laboratory flumes to study the
bed form structures, sediment suspension and the influence of bed roughness over flat
sediment beds of different composition \cite{other}.

In the present work we study   the sediment bed topography and geometry of bed
forms due to the the effect of upstream
propagating waves against the flow. In particular we will concentrate on the nature of the bedform near critical frequency regime when  wave blocking
is in force on water surface.

In order to observe the sediment bed form structures
 a sand bed of thickness  10cm and of length  8m covering the entire
width (50cm) of the flume is laid at the bottom of the flume. The median particle diameter
of the sand is 0.25mm with a standard geometric deviation $\sigma_g $ = 0.685. The specific gravity
of sediments used for the experiments is 2.65.

Series of experiments were conducted over the
plane sediment bed to examine the effect of upstream propagating waves against the flow on
the bed form structures that usually develop along the flow as regular ripples. In the first set of experiments we  monitor the bed form structures only using the unidirectional
flow of a given discharge over the plane sediment bed for a certain period of time to observe
the uniform sand waves on the bed surface without the obstacle. In the second set of experiments we examine the bed profile when the   counter propagating waves, (generated from the wavemaker), of fixed  frequencies are superimposed on the flow. In particular we adjust the frequency so that the blocking condition is reached for a specific discharge and a horizon is formed. We repeat the process for different discharges and in each case we note the ripple pattern in the sediment throughout the sand bed when the blocking condition is reached. In all the cases  a certain time, when the bed form gets nearly equilibrium
state, the flow velocity is stopped and subsequently the bed elevations are measured at an interval of  every 2cm
 from upstream to downstream along the center line of the flume covering the distance
of about 3.5m. The instrument we use is  Micro-acoustic
Doppler velocimeter (ADV) made by Sontek, USA. Similar process is repeated to examine the effect of white hole horizon on the sediment bed
form structures for different pairs of discharge and frequency. In Figures 10a and 10b we plot the exact (real space) ripple heights against distance for two different discharges, ie. $ 0.018 m^3/sec $ and $0.019 m^3/sec $.  The bed elevation data analysis from Figures 10a and 10b
suggests that there is a significant change in the bed form structures due to the wave blocking
where the amplitude of upstream propagating waves is almost zero value. Our findings are summarized in Table 2.

\vspace {.3cm}
{\bf{Section VI: Conclusions  and future prospects}}\\
\vspace {.5cm}

In general, it is widely accepted that in a uniform open channel flow, when
dimensionless shear stress exceeds the critical shear stress value the initiation of sediment
motion starts on the flat bed surface. As the fluid velocity increases the bed shear stress increases and sediment transport starts where sediment is removed and transported locally.  After a certain time it is observed that the sediment
bed is changed completely forming a kind of regular asymmetric sand waves like ripples
following the direction of the flow. As fluid flows over a flat sediment bed in a flume, due to
the excess shear stress the bed forms are generated along the flume surface. The sediment is transported from the stoss-side and deposited in the lee-side of the ripple. The size and shape of ripples are uniform in a statistical sense and depends on the flow and bed features.  Generically the ripples move downstream along the flow much slowly than the fluid. Indeed the sediment bed and fluid becomes a strongly coupled system with increase in flow velocity and results in an extremely complicated dynamical system even for a uniform fluid flow. There are various analytical models that attempt to predict the nature and motion of the ripples for a given flow but there are many important issues where consensus is not reached. (See \cite{dey} for an exhaustive study.) In our case, on top of the steady flow, there is the impressed gravity waves and to the best of our knowledge there is not much literature both in the form of  experimental results or theoretical analysis. Some works on sediment transport have appeared   where the impressed waves and the uniform flow both are in the same direction \cite{maz} but probably not much when the waves and uniform flow are in opposite direction.

Let us summarize our findings.\\
Our work can be divided in to two parts: (i) experiment with obstacle and without sediment and (ii) experiment without obstacle and with sediment.\\
(i) The first part of our work essentially reproduces early observations by \cite{wein,rou1}. However our flume is of a considerably larger dimension than that used in \cite{wein} hence within the restrictions present in our instrumentation and laboratory setup it is not possible for us to reach the level of accuracy that is need to reproduce the experimental results found in \cite{wein}. On the other hand our flume is comparable in size with that of \cite{rou1} so we have been able to derive quantitative estimates of the critical frequencies (of the imposed waves) for different discharges above which frequencies the waves are effectively blocked. The findings are summarized in Table 1. Three discharges are considered and for each one a blocking zone appears on top of the obstacle beyond which the wave amplitude decreases abruptly and effectively vanishes. Furthermore as the discharge increases the critical frequency value decreases showing that, as expected, the higher frequencies are blocked more easily and higher discharge is required to block lower frequencies.\\
(ii) In the second part of our work we use a distinctly different setup where the obstacle is replaced by  a movable sediment bed consisting of sand. Our aim is to study sediment transport in presence of the horizon formation and blocking phenomenon. Specifically we have studied the ripple pattern on the sand bed (Figure 7) in detail after waiting for a sufficiently long time, (approximately four hours), to allow the sand bed to reach a  state such  that it is stable over time scales much larger than the time period of the waves.  Comparing with the well-known  ripple pattern in conventional uniform flows (without wave), Figure 5, we observe that in the present case the sediment bed is divided in to three sectors: in the flow dominated  region (far away from the blocking zone but closer to the discharge source) the ripple structure for uniform flow prevails. The ripple structure is statistically uniform with asymmetrical shape ie. the ripples are sharper in the wave side and slopes more gently along the flow direction. Below the analogue white hole dominated blocking zone the sand ripples are more disorganized and on the average of lower height. Again  in the wave-dominated region (far away from the horizon but closer to the wavemaker) the ripples are more symmetrical due to the interaction between the flow and wave force that tend to destroy the asymmetrical nature of the ripples and induces a more symmetrical shape. Therefore, the exploratory data analysis suggests that there is a
 transition  in the bed form pattern in the flume with asymmetric ripples induced by the flow, followed by  flat
 small sand bars in the blocking region and more symmetrical ripples in the wave dominated region. This suggests that the sediment bed  may be segmented into three regions such as, the flow region at the upstream, blocking
 region, i.e. near  analogue white hole horizon, and the wave region at the downstream respectively. Decreasing values of the standard deviation near the blocking condition in Table 2 also corroborates this.

 Let us put our work in its proper perspective as it can have relevance in two different disciplines: On the one hand, in classical fluid dynamics it can be applied in practical situations having wave blocking zones where sediment bed profile plays an important role. If the analytic theory of sediment transport in the presence of opposing flow and wave is sufficiently developed its predictions can be tested in experimental setups like ours.  On the other hand, in analogue gravity scenario, it can serve as an analogue of a different form of matter  that couples with background gravity.

 As a future work we wish to improve the statistics by making further observations with different discharge and frequencies. Another experimental aspect would be to study the turbulence phenomenon from the velocity spectrum of the fluid and sediment particles. Work is in progress along these directions.

\vspace {.3cm}
{\bf{Acknowledgements:}} It is indeed a pleasure to thank Silke Weinfurtner for early suggestions regarding the experimental setup. We are grateful to Koustuv Debnath  and Satya Praksh Ojha for helpful discussions. The work of P. Das is funded by INSPIRE, DST, India.

\begin{table}[htbp]
	\centering
	\caption{}
	\begin{tabular}{|r|r|r|r|r|r|}
		\hline
		\multicolumn{6}{|c|}{Flow Discharge} \\
		\hline
		\multicolumn{2}{|c|}{0.0141$m^3/sec$} & \multicolumn{2}{|c|}{0.0166$m^3/sec$} & \multicolumn{2}{|c|}{0.0194$m^3/sec$} \\
		\hline
		Frequency/sec & Amplitude(cm) & Frequency/sec & Amplitude(cm) & Frequency/sec & Amplitude(cm) \\
		\hline
		\multicolumn{1}{|c|}{1.35} & \multicolumn{1}{|c|}{7} & \multicolumn{1}{|c|}{0.83} & \multicolumn{1}{|c|}{8} & \multicolumn{1}{|c|}{0.62} & \multicolumn{1}{|c|}{11} \\
		\hline
		\multicolumn{1}{|c|}{1.67} & \multicolumn{1}{|c|}{2} & \multicolumn{1}{|c|}{1.06} & \multicolumn{1}{|c|}{7} & \multicolumn{1}{|c|}{0.83} & \multicolumn{1}{|c|}{10} \\
		\hline
		\multicolumn{1}{|c|}{1.9} & \multicolumn{1}{|c|}{1} & \multicolumn{1}{|c|}{1.26} & \multicolumn{1}{|c|}{1} & \multicolumn{1}{|c|}{1.06} & \multicolumn{1}{|c|}{9} \\
		\hline
		\multicolumn{1}{|c|}{2.06} & \multicolumn{1}{|c|}{0} & \multicolumn{1}{|c|}{1.35} & \multicolumn{1}{|c|}{-1} & \multicolumn{1}{|c|}{1.23} & \multicolumn{1}{|c|}{3} \\
		\hline
		\multicolumn{1}{|c|}{2.26(blocking)} & \multicolumn{1}{|c|}{0} & \multicolumn{1}{|c|}{1.6(blocking)} & \multicolumn{1}{|c|}{-2} & \multicolumn{1}{|c|}{1.4(blocking)} & \multicolumn{1}{|c|}{-1} \\
		\hline
	\end{tabular}%
	\label{tab:addlabel}%
\end{table}%

\begin{table}[htbp]
	\centering
	\caption{}
	\footnotesize\setlength{\tabcolsep}{-0.2pt}
	\begin{tabular}{|c|c|c|c|c|c|c|c|c|}
		\hline
		\multicolumn{7}{|c|}{Actual Ripple Height in blocking condition measured from smooth bed} & \multicolumn{2}{c|}{Same in only uniform flow} \\
		\hline
		Flow($m^3/sec$) & \multicolumn{2}{c}{Blocking Position} & \multicolumn{2}{|c|}{Towards Flow} & \multicolumn{2}{|c|}{Towards Wave Maker} &       &  \\
		\hline
		& Mean & Standard & Mean & Standard & Mean & Standard & Mean & Standard \\
		& Height(cm) & Deviation(cm) & Height(cm) & Deviation(cm) & Height(cm)  & Deviation(cm) & Height(cm) & Deviation(cm) \\
		
		\hline
		0.0208 & 6.3269 & 0.5205 & 7.1064 & 1.348 & 6.7621 & 0.5744 & 6.8091 & 1.0962 \\
		\hline
		0.0197 & 6.8413 & 0.769 & 7.3256 & 1.7491 & 6.8365 & 0.5425 & 8.301 & 0.8419 \\
		\hline
		0.0183 & 6.4794 & 0.4249 & 6.856 & 1.4154 & 7.01  & 1.2608 & 8.0356 & 0.7111 \\
		\hline
		0.0169 & 7.3173 & 0.5381 & 7.495 & 0.5536 & 7.2564 & 0.6033 & 8.5549 & 1.1738 \\
		\hline
	\end{tabular}%
	\label{tab:addlabel}%
\end{table}

\begin{figure}
	{\centerline{\includegraphics[width=12cm, height=10cm] {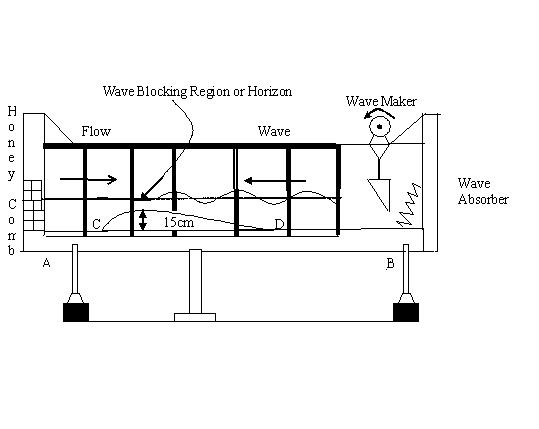}}}
	\caption{Schematic diagram of the flume with obstacle at blocking condition. AB = 10m is the straight portion of the flume. CD = 2.28m is the obstacle.} \label{fig1}
\end{figure}

\begin{figure}
	{\centerline{\includegraphics[width=12cm, height=8cm] {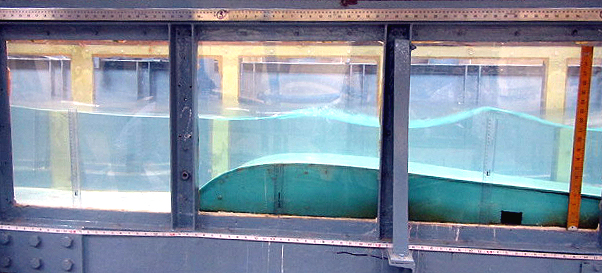}}}
	\caption{Wave blocking with obstacle in our flume} \label{fig2}
\end{figure}

\begin{figure}
	{\centerline{\includegraphics[width=12cm, height=8cm] {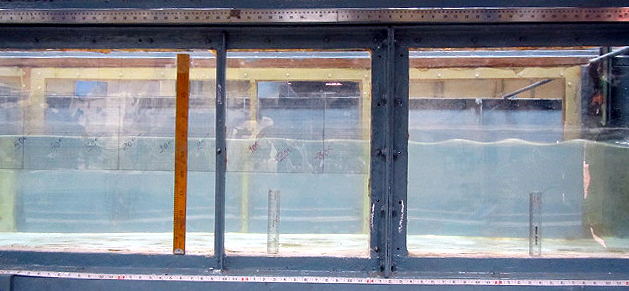}}}
	\caption{Wave blocking with no obstacle} \label{fig3}
\end{figure}

\begin{figure}
	{\centerline{\includegraphics[width=12cm, height=8cm] {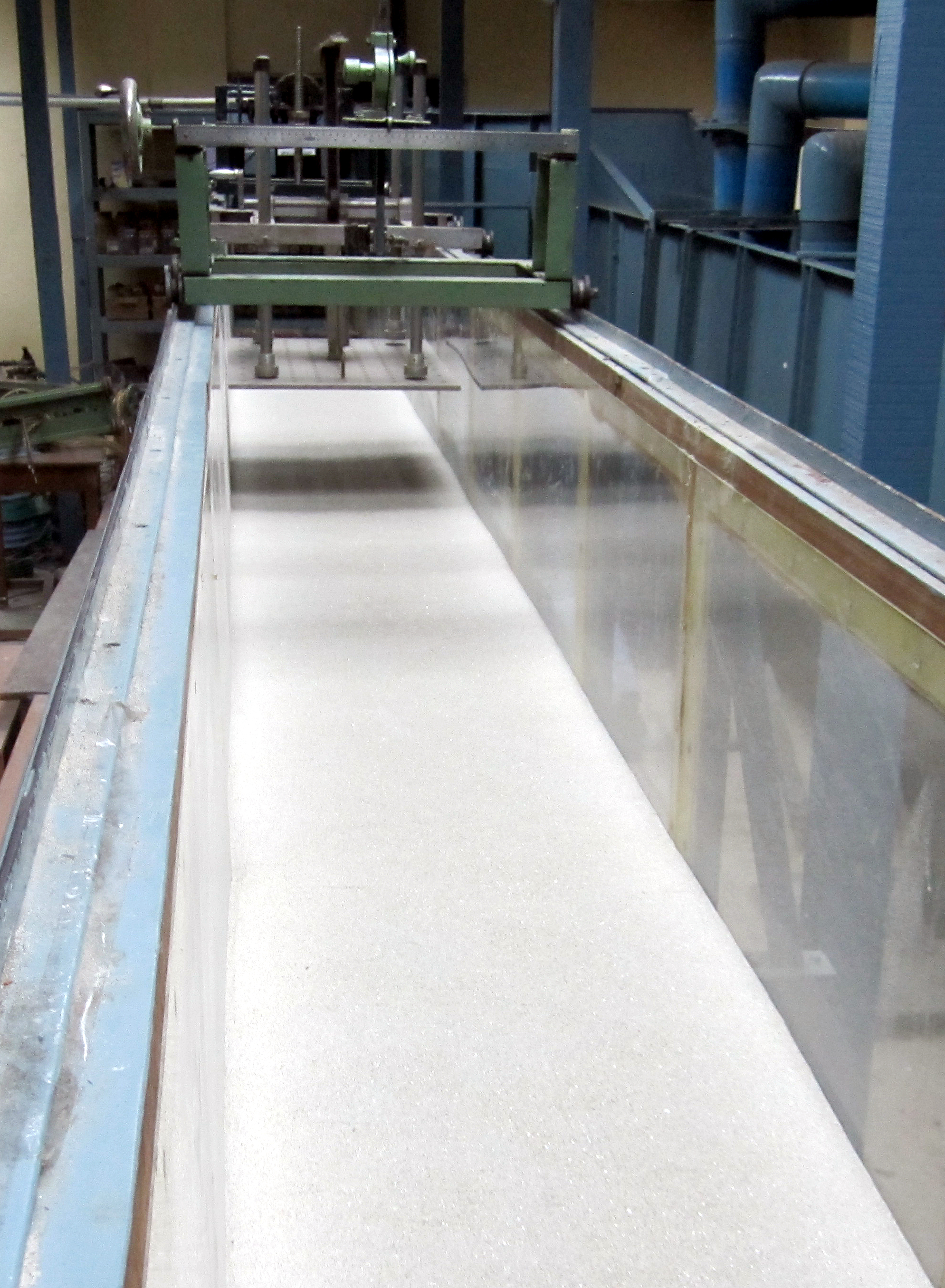}}}
	\caption{Flume with smooth sediment bed before experiment.} \label{fig4}
\end{figure}

\begin{figure}
	{\centerline{\includegraphics[width=12cm, height=8cm]{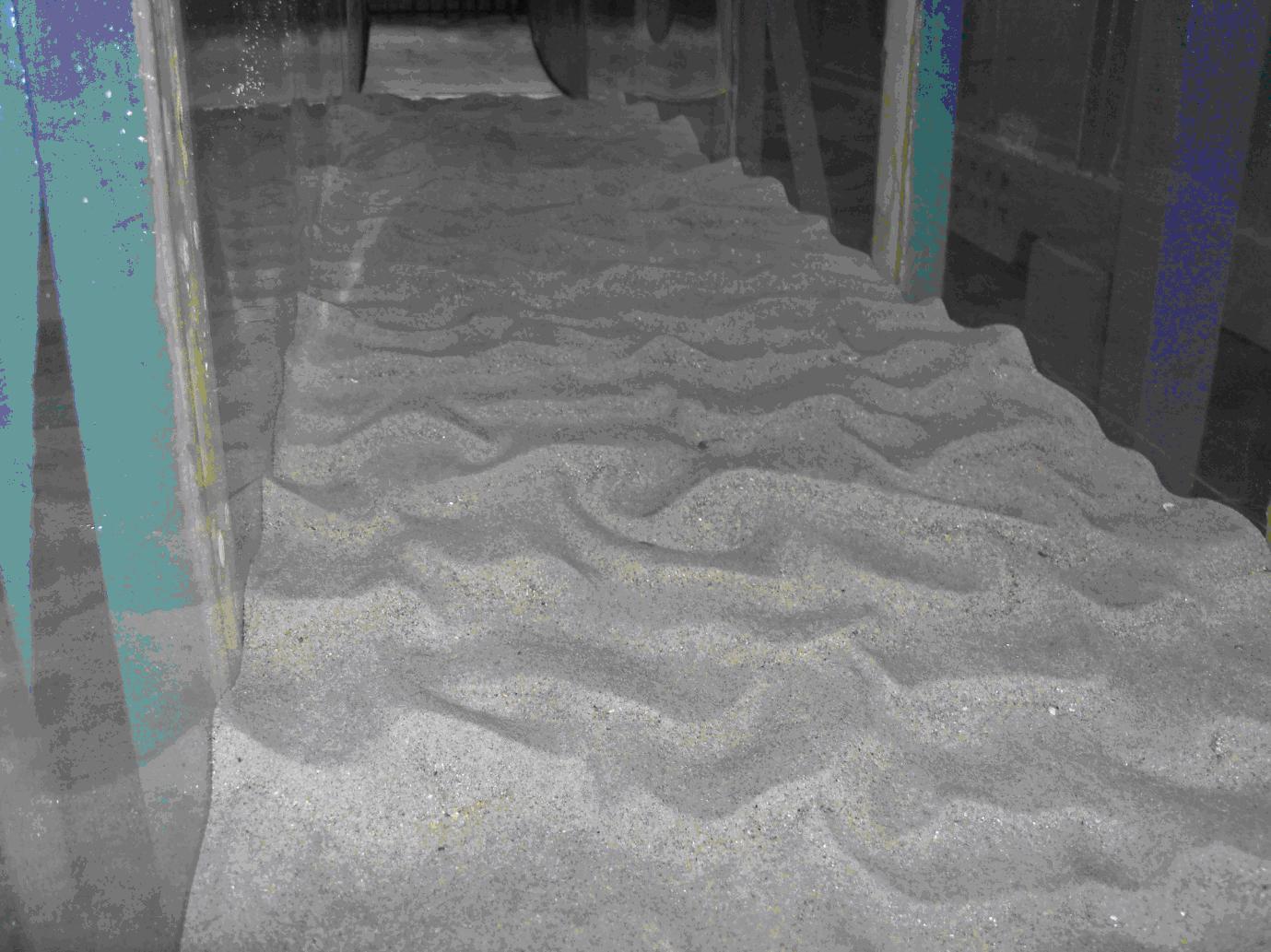}}}
	\caption{Ripples with only flow} \label{fig5}
\end{figure}

\begin{figure}
	{\centerline{\includegraphics[width=12cm, height=8cm]{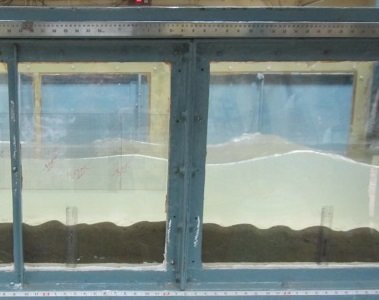}}}
	\caption{Wave blocking over the sediment bed} \label{fig6}
\end{figure}

\begin{figure}
	{\centerline{\includegraphics[width=12cm, height=8cm]{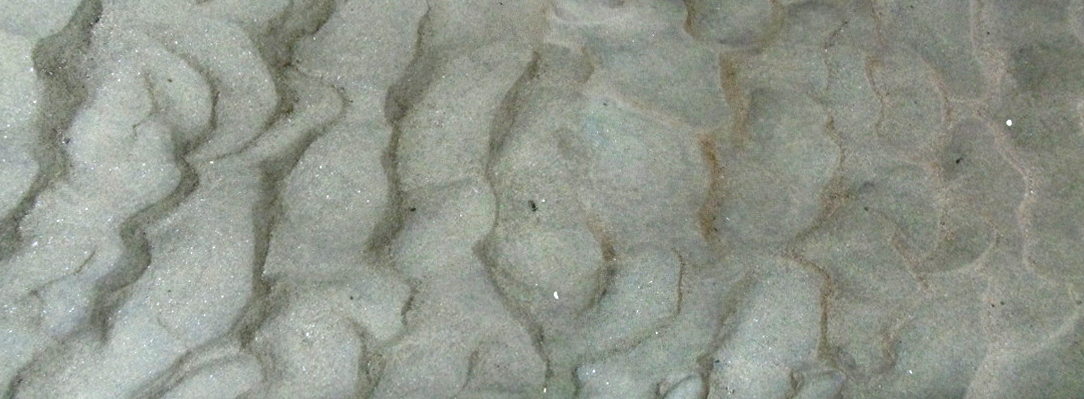}}}
	\caption{Effect of analogue White Hole horizon on sediment bed. The ripples are higher and more regular on the left side near the flow dominated region, similar to ripples with only uniform flow. Ripples on the right are small flat bar-like. The latter are below the horizon dominated region. } \label{fig7}
\end{figure}
\begin{figure}
	{\centerline{\includegraphics[width=12cm, height=8cm] {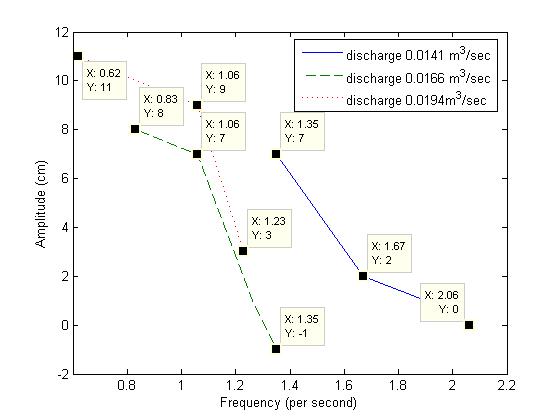}}}
	\caption{Amplitude vs. frequency plot for three different discharges. The value of  frequency, (above which the amplitude abruptly vanishes), is called the critical frequency for that particular discharge. Here X and Y correspond to X-axis and Y-axis respectively.} \label{fig8}
\end{figure}

\begin{figure}
	{\centerline{\includegraphics[width=12cm, height=8cm] {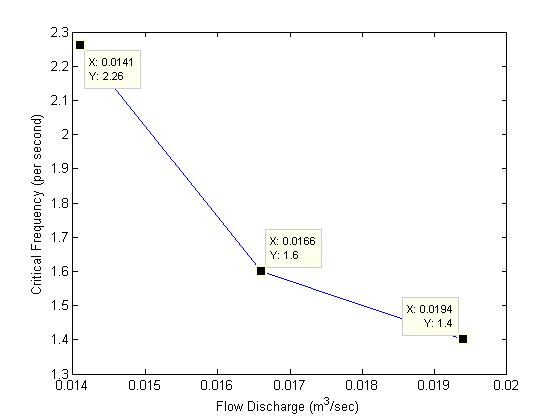}}}
	\caption{Critical frequency vs. discharge plot. The value of critical frequency decreases as the discharge increases. Here X and Y correspond to X-axis and Y-axis respectively.}  \label{fig9}
\end{figure}

\begin{figure}
	\centering
	\subfigure[]{\includegraphics[width=12cm, height=7cm]{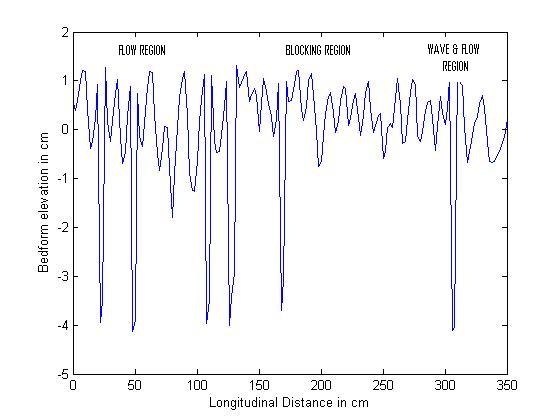}}
	\hfill
	\subfigure[]{\includegraphics[width=12cm,height=7cm]{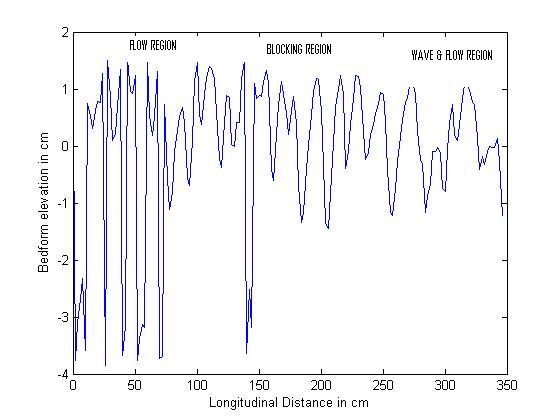}}
	\hfill
	\caption{Actual ripple height vs. distance measured from A, as shown in  Figure 1. Clearly the ripples on the left, near the flow dominated region, are higher than the ripples on the right near the horizon dominated region.}\label{fig10}
\end{figure}

\end{document}